\documentclass{emulateapj}

\newcommand{\f}{\phantom{2}}

\newcommand{\gtsimeq}{\raisebox{-0.6ex}{$\,\stackrel 
        {\raisebox{-.2ex}{$\textstyle >$}}{\sim}\,$}} 
\newcommand{\umu}{\mu}

\newcommand{\civ}{C\,{\sc iv}}
\newcommand{\lya}{Ly\,$\alpha$}

\newcommand{\hbeta}{H\,$\beta$}
\newcommand{\halpha}{H\,$\alpha$}
\newcommand{\oiii}{[O\,{\sc iii}]}

\newcommand{\sii}{[S\,{\sc ii}]}
\newcommand{\myemail}{chris.willott@nrc.ca}
\def\chandra{{\it Chandra~}}
\def\xmm{{\it XMM-Newton~}}

\shorttitle{Dust and gas obscuration in EDXS reddened quasars}
\shortauthors{Willott et al.}

\begin{document}


\title{Dust and gas obscuration in ELAIS Deep X-ray Survey reddened quasars}


\author{Chris J.\ Willott} 
\affil{Herzberg Institute of Astrophysics, National Research Council,
5071 West Saanich Rd,\\ Victoria, B.C. V9E 2E7, Canada\\ {\tt email:
\myemail}}
\author{Chris\ Simpson}
\affil{Department of Physics, University of Durham, South Road, Durham DH1 3LE, UK}
\author{Omar\ Almaini}
\affil{Institute for Astronomy, University of Edinburgh, Royal
Observatory,  Blackford Hill, Edinburgh EH9 3HJ, UK}
\author{Olivia\ Johnson}
\affil{Institute for Astronomy, University of Edinburgh, Royal
Observatory,  Blackford Hill, Edinburgh EH9 3HJ, UK}
\author{Andrew\ Lawrence}
\affil{Institute for Astronomy, University of Edinburgh, Royal
Observatory,  Blackford Hill, Edinburgh EH9 3HJ, UK}
\author{James S.\ Dunlop}
\affil{Institute for Astronomy, University of Edinburgh, Royal
Observatory,  Blackford Hill, Edinburgh EH9 3HJ, UK}
\author{Nathan D.\ Roche}
\affil{Institute for Astronomy, University of Edinburgh, Royal
Observatory,  Blackford Hill, Edinburgh EH9 3HJ, UK}
\author{Robert G.\ Mann}
\affil{Institute for Astronomy, University of Edinburgh, Royal
Observatory,  Blackford Hill, Edinburgh EH9 3HJ, UK}
\author{James C.\ Manners} 
\affil{Dipartimento di Astronomia, dell'Universita di Padova, Vicolo
dell'Osservatorio, 2-35122 Padova, Italy}
\author{Eduardo\ Gonz\'alez-Solares}
\affil{University of Cambridge, Institute of Astronomy, The Observatories, Madingley Road, Cambridge, CB3 0HA, UK}
\author{Ismael\ P\'erez-Fournon}
\affil{Instituto de Astrof\'\i sica de Canarias, C/ Via
Lactea s/n, 38200 La Laguna, Tenerife, Spain}
\author{Rob J.\ Ivison}
\affil{UK Astronomy Technology Centre, Royal Observatory, Blackford Hill, Edinburgh, EH9 3HJ, UK}
\author{Stephen\ Serjeant}
\affil{Centre for Astrophysics and Planetary Science, School of
Physical Sciences, University of Kent, Canterbury, Kent, CT2 7NZ, UK}
\author{Seb J.\ Oliver}
\affil{Astronomy Centre, Department of Physics \& Astronomy, University of Sussex, Falmer, Brighton, BN1 9QH, UK}
\author{Richard G.\ McMahon}
\affil{University of Cambridge, Institute of Astronomy, The Observatories, Madingley Road, Cambridge, CB3 0HA, UK}
\author{Michael\ Rowan-Robinson}
\affil{Astrophysics Group, Blackett Laboratory, Imperial College,
Prince Consort Rd., London, SW7 2BW UK}

\begin{abstract}

Hard X-ray surveys have uncovered a large population of heavily
obscured AGN. They also reveal a population of quasars with moderate
obscuration at both visible and X-ray wavelengths. We use \chandra
selected samples of quasars from the ELAIS Deep X-ray Survey (EDXS)
and the \chandra Deep Field-North to investigate the obscuration
towards the nuclei of moderately obscured AGN. We find an inverse
correlation between the optical to X-ray flux ratio and the X-ray
hardness ratio which can be interpreted as due to obscuration at
visible and X-ray wavelengths. We present detailed optical and
near-infrared data for a sample of optically-faint ($R>23$) quasars
from the EDXS. These are used to constrain the amount of rest-frame
UV/optical reddening towards these quasars. It is found that
optically-faint quasars are mostly faint due to obscuration, not
because they are intrinsically weak. After correcting for reddening,
the optical magnitudes of most of these quasars are similar to the
brighter quasars at these X-ray fluxes. Combining with gas column
densities inferred from the X-ray observations we consider the
gas-to-dust ratios of the obscuring matter. We find that the quasars
generally have higher gas-to-dust absorption than that seen in the
Milky Way -- similar to what has been found for nearby Seyfert
galaxies. We consider the possible existence of a large population of
X-ray sources which have optical properties of Type 1 (unobscured)
quasars, but X-ray properties of Type 2 (obscured) quasars. However,
we find that such sources only contribute about 6\% of the 0.5-8 keV
X-ray background. Finally we show that the observed distribution of
optical-to-X-ray flux ratios of quasars at $z>1$ is skewed to low
values compared to the intrinsic distribution due to the fact that the
observed-frame $R$-band light is emitted in the UV and is more easily
obscured than hard X-rays.

\end{abstract}


\section{Introduction}

The hard spectral shape of the X-ray background led to the idea that a
large population of obscured active galactic nuclei (AGN) exist which
fail to show up in optically-selected quasar surveys (Comastri et
al. 1995).  Deep hard X-ray surveys with \chandra and \xmm have now
revealed the sources responsible for the 0.5-10 keV X-ray background
(e.g. Hornschemeier et al. 2001; Tozzi et al. 2001; Hasinger et
al. 2001; Manners et al. 2003). As expected, the majority of the
optical counterparts to hard X-ray sources are galaxies containing
optically-obscured AGN.

Seyfert galaxies at low redshift often show complex absorption
structures consisting of both cold (neutral) and warm (partially
ionized) absorbers (Mushotzky, Done \& Pounds 1993). A fraction of
Seyferts appear to be Compton thick and their observed X-ray emission
is reflection dominated. Due to the low signal-to-noise of the X-ray
spectra of the sources responsible for the X-ray background, there is
only limited knowledge of their absorption properties. It is unknown
whether the orientation-based obscuration scheme which works well for
low-redshift Seyferts can also be applied at higher redshifts. There
are still only a handful of Type 2 narrow-line quasars found in deep
X-ray surveys (Almaini et al. 1995; Norman et al. 2002; Stern et
al. 2002; Crawford et al. 2002; Mainieri et al. 2002; Szokoly et
al. 2004), seemingly at odds with simple unification schemes (however
there are still many faint objects without redshifts in these surveys
which could be Type 2 quasars). The relative lack of luminous Type 2
quasars and the dominance of obscured AGN at lower luminosities (Ueda
et al. 2003; Hasinger et al. 2003) suggests that the fraction of
obscured objects is a strong function of luminosity, as had previously
been found for low-radio-frequency selected AGN (Simpson, Rawlings \&
Lacy 1999; Willott et al. 2000). The fraction of Compton thick AGN in
the X-ray background sources is also quite uncertain (e.g. Fabian,
Wilman \& Crawford 2002).

The dust properties of the high redshift X-ray absorbing material have
not yet been well studied. There are certainly objects at low redshift
with dust absorption quite different from that expected from the
observed X-ray absorption based on a Galactic gas-to-dust ratio
(Simpson 1998; Maiolini et al. 2001a) and a few cases of such
discrepancies at higher redshifts have been reported (Akiyama et
al. 2000; Risaliti et al. 2001; Willott et al. 2003; Watanabe et
al. 2004). In this paper we discuss the dust and X-ray absorption
present in hard X-ray sources contained within the ELAIS Deep X-ray
Survey and the implications for understanding the sources responsible
for the X-ray background.

In Sec.\,2 we discuss the observed correlations between optical and
X-ray properties of X-ray selected quasars and what these suggest
about obscuration. Sec.\,3 presents near-infrared and optical data for
optically-faint X-ray selected quasars. In Sec.\,4 we fit model quasar
spectra to the observations to constrain the reddening towards these
quasars. In Sec.\,5 we compare the obscuration in the UV/optical with
that in X-rays to determine the gas-to-dust ratio of the obscuring
material and compare this with values in the Milky Way, the Small
Magellanic Cloud and low-redshift AGN. In Sec.\,6 we discuss the
effect that obscuration plays in altering the intrinsic
optical-to-X-ray flux ratios of quasars to those observed.  We assume
throughout that $H_0=70~ {\rm km~s^{-1}Mpc^{-1}}$, $\Omega_{\mathrm
M}=0.3$ and $\Omega_\Lambda=0.7$.

\section{Optically faint quasars in \chandra surveys}
\label{optfaint}

In this paper, we will concentrate on X-ray sources which are
associated with unresolved optical counterparts. With the exceptions
of a few stars (readily identified by their high optical to X-ray flux
ratios) these sources have optical and (usually) near-IR emission
which is dominated by the active nucleus. We will commonly refer to
these objects as quasars for simplicity, but the reader should bear in
mind that the moderate optical and X-ray luminosities place them
around the border with Seyfert Type 1s. It may seem strange that we
decide to concentrate on quasars when galaxies dominate the hard X-ray
background sources. However, quasars likely dominate the
high-redshift, moderate-luminosity population (Barger et al. 2002;
Hasinger et al. 2003; Steffen et al. 2003) and the presence of optical
AGN continuum emission allows a more detailed investigation of the
optical obscuration toward the nucleus, unlike the totally
optically-obscured X-ray sources associated with galaxies.

The ELAIS Deep X-Ray Survey (EDXS) consists of two 75 ks \chandra
pointings in northern ELAIS (European Large Area ISO Survey; Oliver et
al. 2000) fields which have detected a total of 233 X-ray
sources. Full details of the EDXS are given in Manners et al. (2003)
and Gonzalez-Solares et al. (2004). In this paper we only consider
sources in the EDXS-N2 region at RA=$16^{\rm h}36^{\rm m}47^{\rm s}$
DEC=$+41^{\rm d}01^{\rm m}34^{\rm s}$ (J2000.0), because this region
has superior optical imaging (seeing 0.75 arcsec) which allows the
separation of unresolved point-sources and extended galaxies (which
have typical sizes $ \gtsimeq 1$ arcsec) to a magnitude limit of
$R=25$. As we will see below, classification at this magnitude limit
is essential for identifying quasars at faint X-ray fluxes.

To increase the number of data points and therefore decrease
statistical uncertainties in some of our analyses, we will also use
X-ray sources from the \chandra Deep Field-North Survey (CDF-N). The
1Ms \chandra exposure detected 370 point sources (Brandt et
al. 2001). Deep optical imaging of this field has been obtained and,
crucially for our analyses, optically-compact sources have been
identified and flagged `C' for compact (Barger et al. 2002). Sources
showing broad emission lines in their optical spectra are flagged `B'
and sources both compact and with broad lines `BC'. We include as
quasars sources flagged as `BC' or `C' and also the four sources
flagged `B' at redshifts $z>0.8$ since the optical luminosities of
these suggest they are dominated by an AGN and the fact they were not
classified as compact is likely due to the proximity of companions on
the images.


\begin{figure}
\resizebox{0.48\textwidth}{!}{\includegraphics{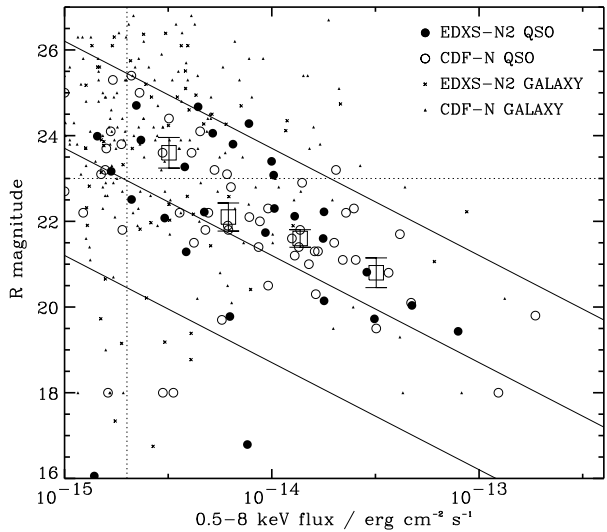}}
\caption{ Optical magnitude versus full band X-ray flux for sources in
the EDXS-N2 and CDF-N surveys. Quasars (as defined in Sec.\ 2) are
shown with circles. There is a well-known correlation between the
optical and X-ray flux which is visible here for the quasars. The
solid lines show constant optical to X-ray flux ratios of 0.1, 1 and
10 (top to bottom). The `quasars' with very high optical to X-ray flux
ratios ($>10$) in the bottom-left corner are stars. The vertical
dotted line shows the X-ray flux-limit considered throughout most of
this paper. Quasars above this flux limit have been binned in X-ray
flux and the symbols with error bars show the median $R$ magnitude
against median X-ray flux in each bin with the associated standard
error. The horizontal dashed line is at $R=23$ and quasars above this
line are classified as optically-faint.
\label{fig:optxflux}}
\end{figure}

\subsection{Correlation between optical and X-ray fluxes}

Fig.\,\ref{fig:optxflux} shows the $R$-band magnitude versus X-ray
flux for X-ray sources from the EDXS-N2 and CDF-N surveys. Sources are
shown as quasars if they are optically-compact in EDXS-N2 or are
quasars as defined above in the CDF-N.  The vertical line shows a
full-band (0.5-8 keV) flux limit of $>2 \times
10^{-15}$\,erg\,cm$^{-2}$\,s$^{-1}$ which will be adopted for the
remainder of this paper. This flux limit was chosen since it is
equivalent to a signal-to-noise of $5\sigma$ for most of the EDXS-N2
\chandra image. We use the full band flux rather than just the hard
band (2-8 keV) flux since the energy dependent response of \chandra is
such that even moderately hard sources may be detectable in the full
band, but not in the hard band. Note that at this flux limit a quasar
with an optical to X-ray flux ratio (the ratio between the fluxes in
the $R$-band and the 0.5-8 keV band) of 0.1 has $R=25.5$ -- close to
the limiting magnitude at which one can separate unresolved from
extended sources in ground-based optical imaging.

\begin{figure}
\vspace{0.8cm}
\resizebox{0.48\textwidth}{!}{\includegraphics{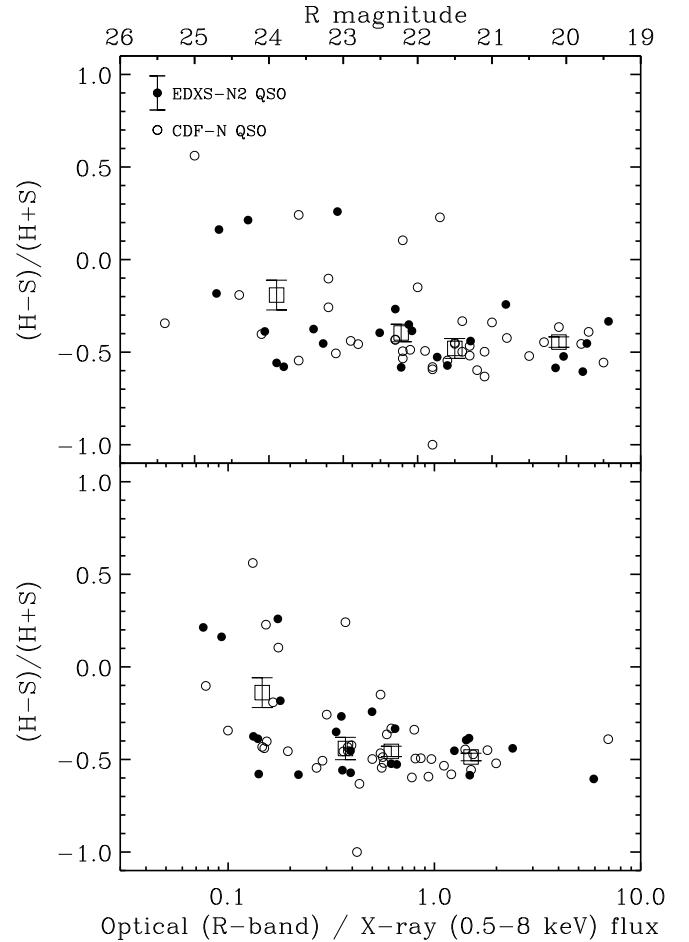}}
\vspace{0.6cm}
\caption{{\it Upper:} Hardness of X-ray emission against $R$-band
magnitude for quasars above a 0.5-8 keV flux limit of $>2 \times
10^{-15}$\,erg\,cm$^{-2}$\,s$^{-1}$ and with X-ray signal-to-ratio $>
5$. The mean error bar on the hardness of these data is shown on the
EDXS-N2 symbol label. The open squares show the mean hardness binned in
optical magnitude and the associated standard error. {\it Lower:}
Hardness of X-ray emission against optical to X-ray flux ratio for the
same quasars as in the upper panel. The open squares show the mean
hardness binned in optical to X-ray flux ratio and the associated
standard error. 
\label{fig:hard}}
\end{figure}


Fig.\,\ref{fig:optxflux} shows that the quasars tend to have a fairly
small range of optical to X-ray flux ratio as has been demonstrated
before (Schmidt et al. 1998). There is some indication that the
typical optical to X-ray flux ratio for quasars decreases at faint
X-ray fluxes (in the faintest bin on Fig.\,\ref{fig:optxflux}) but
this is not statistically significant. However, if there are
significant numbers of $R>25$ quasars which we have not identified as
such, this difference and its significance could be increased. We will
investigate in this paper how this plot of optical and X-ray fluxes is
affected by the existence of a sizeable number of partially obscured
quasars.

\subsection{The X-ray hardness ratio and optical and X-ray fluxes}
\label{hardratio}

In order to determine whether obscuration may play a role in the
observed optical to X-ray ratios, we now consider the hardness of the
X-ray spectra of these sources as defined by (H-S)/(H+S) (where H is
the flux in the 2-8 keV band and S is the flux in the 0.5-2 keV band;
see Manners et al. 2003 for more details). In the \xmm survey of the
Lockman Hole, Mainieri et al. (2002) showed that the full range of
hardness values in surveys such as these can be explained by X-ray
sources with a constant intrinsic power-law slope of $\Gamma \approx
2$ and a range of obscuring columns of $10^{19} < N_{\rm H} < 10^{24}
~{\rm cm}^{-2}$. Therefore we will use the observed hardness as an
indicator of the gaseous obscuring column since the sources we discuss
typically do not have enough counts for detailed spectral analyses. It
should be remembered that a range of intrinsic power-law slopes could
account for some of the scatter in observed hardness ratios.

Fig.\,\ref{fig:hard} shows the hardness ratio for quasars in the
EDXS-N2 and CDF-N surveys plotted against $R$-band
magnitude and optical to X-ray flux ratio. The binned data show some
evidence that at faint optical magnitudes and at low optical to X-ray
flux ratios, the typical hardness values of quasars increases,
presumably because the X-ray emission becomes more heavily obscured.  To test
whether there are significant correlations we perform a Spearman rank
correlation analysis. The results show that there is a correlation
between $R$-band magnitude and hardness significant at $>99$\%
confidence and between optical to X-ray flux ratio and hardness at
$>99.9$\%. 


\begin{deluxetable*}{lccrccccc}
\tabletypesize{\scriptsize}
\tablecaption{Table of observed objects. \label{tbl-1}}
\tablewidth{0pt}
\tablehead{
\colhead{Source} &                              \colhead{Near-IR position}   & \colhead{$z$} &
\colhead{0.5-8 keV flux} &
\colhead{HR}  & \colhead{$V$ mag} & \colhead{$R$ mag}  & \colhead{$I$ mag} & \colhead{$H$ mag}  
}
\startdata
N2\_15                  & 16:37:04.40 +40:56:24.7 & 1.562 & $\f(2.2  \pm 0.6). 10^{-15}$ & $-0.18 \pm 0.23$ & $25.35 \pm 0.19$ & $24.82 \pm 0.10$  & $24.13 \pm 0.12$ & $20.0  \pm 0.2$ \\
N2\_35\tablenotemark{a} & 16:36:49.28 +41:03:23.7 & 2.741 & $\f(7.8\pm 1.0). 10^{-15}$ & $ \f0.16 \pm 0.31$ & $24.80 \pm 0.09$ & $24.72 \pm 0.07$  & $24.20 \pm 0.10$ & ---             \\
N2\_47                  & 16:36:36.21 +41:05:09.2 & 0.884 & $\f(3.8  \pm 0.7). 10^{-15}$ & $-0.45 \pm 0.18$ & $23.69 \pm 0.10$ & $23.43 \pm 0.05$  & $22.21 \pm 0.05$ & $20.2  \pm 0.2$ \\
N2\_60                  & 16:36:19.18 +41:04:36.5 & 1.498 & $\f(6.5  \pm 1.0). 10^{-15}$ & $-0.58 \pm 0.13$ & $24.35 \pm 0.10$ & $23.82 \pm 0.05$  & $23.04 \pm 0.05$ & $20.5  \pm 0.2$ \\
N2\_61                  & 16:36:18.23 +41:00:37.4 & 1.390 & $\f  (5.2 \pm 0.9). 10^{-15}$& $-0.39 \pm 0.17$ & $25.15 \pm 0.14$ & $24.08 \pm 0.05$  & $23.05 \pm 0.05$ & $20.3  \pm 0.2$ \\
N2\_64                  & 16:36:14.45 +41:03:48.7 & ---   & $(10.0  \pm 1.2). 10^{-15}$  & $-0.38 \pm 0.11$ & $23.30 \pm 0.10$ & $23.00 \pm 0.05$  & $22.24 \pm 0.05$ & $19.5  \pm 0.2$ \\
N2\_68\tablenotemark{a} & 16:37:25.33 +41:00:20.2 & 2.218 & $\f(2.3  \pm 0.6). 10^{-15}$ & $-0.56 \pm 0.11$ & $24.15 \pm 0.07$ & $24.02 \pm 0.05$  & $23.49 \pm 0.08$ & $20.5  \pm 0.2$ \\
 \enddata

\tablenotetext{a}{No near-infrared spectroscopy was obtained for these
sources. Optical spectroscopy is described in Sec.\,\ref{optspec}}

\tablecomments{EDXS-N2 optically-faint quasars with near-infrared or
optical spectroscopy presented in this paper.  0.5-8 keV fluxes are in
units of erg\,cm$^{-2}$\,s$^{-1}$. Hardness ratios are calculated by
HR=(H-S)/(H+S). Optical and near-IR magnitudes are measured in a 3
arcsec aperture. N2\_35 is not detected at $H$-band but is detected at
$K$ with $K=21.04 \pm 0.16$.  The imaging data are described in
Gonz\'alez-Solares et al. (2004), Roche et al. (2002) and Ivison et
al. (2002).}
\end{deluxetable*}


It is important to realize that the $R$-band magnitude is very
strongly anti-correlated with the optical to X-ray flux ratio
($>99.99999$\%) because most sources lie close to the X-ray flux
limit.  Hence the appearance of both variables correlating with
hardness could be induced by just one of the two being physically
related to hardness.  To ascertain if this is true, we employ a
Spearman {\it partial} rank correlation analysis of the correlations
present between the three variables. This method (e.g. Macklin 1982)
assesses the statistical significance of correlations between the two
named variables in the presence of the third. This analysis shows that
the correlation between $R$-band magnitude and hardness (independent
of correlations with optical to X-ray flux ratio) is only significant
at $30$\% confidence. However, the correlation between optical to
X-ray flux ratio and hardness (independent of correlations with
$R$-band magnitude) is statistically significant at $98$\%
confidence. Therefore we find a significant increase in the X-ray
obscuration of quasars at lower optical to X-ray flux ratios. Note
that this correlation has the opposite trend from that expected if the
optical and X-ray obscuration are completely independent.

This increase in X-ray obscuration of quasars at lower optical to
X-ray flux ratios could be explained by an absorbing medium that
contains dust and gas. At the typical redshifts of these quasars
($1<z<3$ -- see Sec.\,\ref{obs}), the observed $R$-band flux is
probing the rest-frame UV and the observed 0.5-8 keV X-rays originate
at harder energies. Therefore the observed $R$-band flux is much more
easily depressed than the X-ray flux. This idea will be revisited in a
more quantitative way in Sec.\,\ref{xoptabs}. First we need to
understand how much optical reddening there is in these quasars and
then compare with the amount of X-ray absorption.

\section{Observations}
\label{obs}

To understand the properties of quasars with low optical to X-ray flux
ratios, we studied all those X-ray sources from the EDXS-N2 which have
optically-unresolved counterparts with $R>23$, full-band fluxes $>2
\times 10^{-15}$\,erg\,cm$^{-2}$\,s$^{-1}$ and \chandra snr $>5$.
There are nine quasars obeying these criteria.
Fig.\,\ref{fig:optxflux} shows that almost all these sources are
located close to the optical to X-ray flux ratio = 0.1 line. One X-ray
source, N2\_25\footnotemark, already has near-infrared and optical
spectroscopy presented in Willott et al. (2003) and is found to be a
quasar with fairly narrow emission lines (2000\,km\,s$^{-1}$) subject
to reddening of $A_V \approx 1$. In contrast to this fairly small
amount of reddening, this quasar has a very hard X-ray spectrum
indicating an obscuring column of $N_{\rm H}= 3 \times 10^{23} \, {\rm
cm}^{-2}$.

\footnotetext{The full IAU name for this source is
CXOEN2~J163655.7+405910. We use the abbreviated names of EDXS-N2
sources in this paper. Cross-referencing of the abbreviated and full
names can be found in Manners et al. (2003)}

\subsection{Near-infrared spectroscopy}
\label{nirspec}

One source in the sample of nine $R>23$ quasars, N2\_35, has a
particularly faint near-infrared counterpart ($K=21$) and hence was
not targeted since it was unlikely to yield a spectrum with
sufficient snr for analysis. The remaining seven quasars apart from
N2\_25 and N2\_35 were the subject of a program of near-infrared
spectroscopy with the OH-airglow suppression Spectrograph (OHS;
Iwamuro et al. 2001) on Subaru Telescope. The observations were
carried out on the nights of UT 2002 May 20-24.  Due to time
constraints we were only able to observe five of our targets. N2\_37
and N2\_68 were not observed. Optical spectra of N2\_68 and N2\_35
have been obtained instead (see Sec.\,\ref{optspec}). The seven quasars which
have data presented in this paper are listed in Table 1. 

The OHS spectrograph uses a fixed grism giving simultaneous $J$ and
$H$-band spectra over the wavelength ranges $1.108-1.353\,\umu$m and
$1.477-1.804\,\umu$m. The seeing was in the range $0.5-0.6$ arcsec for
all observations and a 0.5 arcsec slit was used giving a resolution of
$\approx 40$\,\AA~ (equivalent to $750$\,km\,s$^{-1}$). The total
exposure times per object varied from 3200 - 7000\,s depending upon
their near-IR magnitudes. These integrations were split into nodded
frames of typically 1000\,s each. Stars of spectral type F were
observed immediately after each target to enable correction for
atmospheric extinction.

The OHS data were reduced in a broadly standard manner, with extra
care being taken to remove detector artifacts due to the faintness of
our targets. Wavelength calibration was performed by observing the
planetary nebula NGC 7027, while flux calibration and atmospheric
extinction corrections were determined using the observations of HIP
80419. Further details can be found in Simpson et al. (2004).
Photometric flux-calibration was performed by scaling the reduced
spectra by an aperture correction to account for slit losses.  This
calibration agrees well with the magnitudes from our own near-IR
imaging (Gonz\'alez-Solares et al. 2004).  The snr per resolution
element (4 pixels) in $H$-band for all the reduced spectra lie in the
range 5 - 7.


\begin{figure}
\vspace{-0.1cm}
\resizebox{0.48\textwidth}{!}{\includegraphics{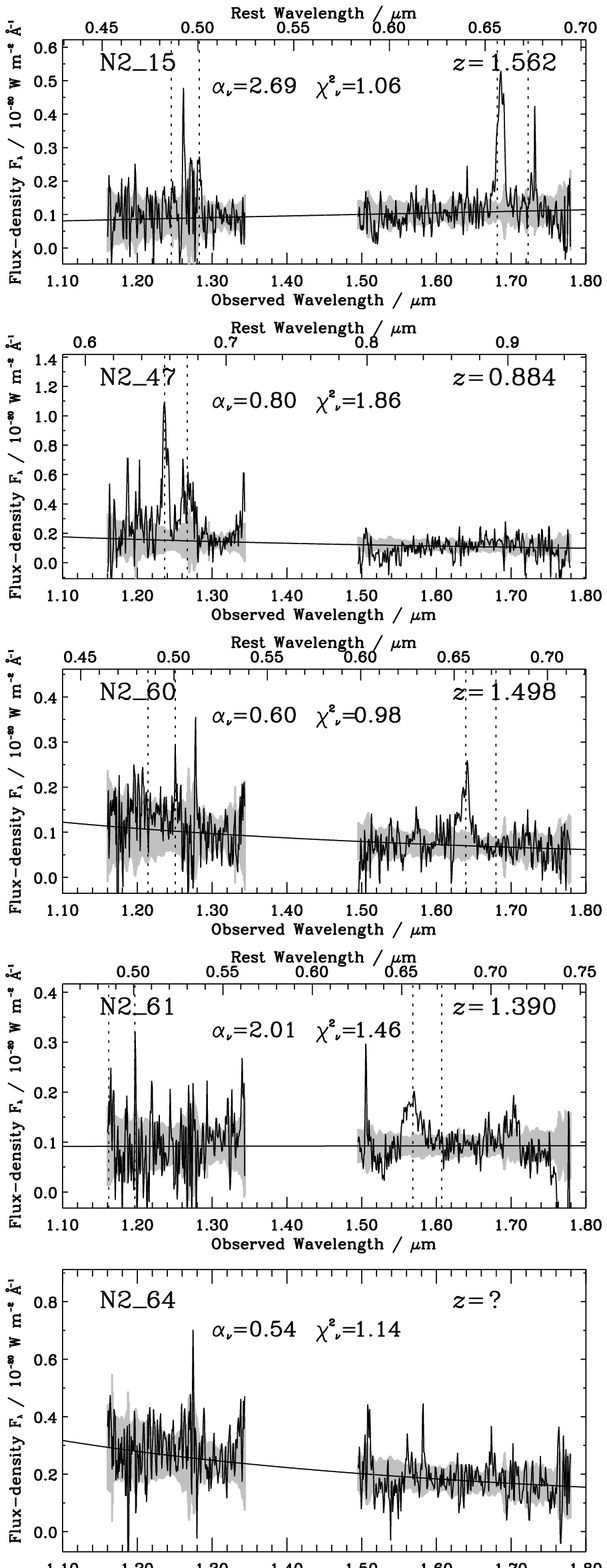}}
\vspace{0.1cm}
\caption{Near-infrared spectra of the five sources observed with the
OHS spectrograph.  Vertical dotted lines mark the expected locations
of the \hbeta\,$\lambda4861$, \oiii\,$\lambda5007$,
\halpha\,$\lambda6563$ and \sii\,$\lambda6716/6731$ emission lines. The
curve is the best fit of a single power-law to the continuum. The
values of the slope $\alpha_{\nu}$ and the goodness-of-fit
$\chi^{2}_{\nu}$ refer to this power-law fit. The grey shaded region
is the $\pm 1 \sigma$ noise added to the best fit power-law
continuum. The gap in the spectra is due to the region between the $J$
and $H$-bands which is heavily affected by atmospheric
absorption. 
\label{fig:spec}}
\end{figure}


\subsection{Optical spectroscopy}
\label{optspec}

For two of the three $R>23$ quasars that we could not obtain
near-infrared spectra of, we have instead obtained optical spectra.
N2\_35 was observed with the GMOS spectrograph at the Gemini-North
Telescope using nod-and-shuffle multi-object spectroscopy for a total
of 3 hours. The data were reduced in a similar manner to as described
in Abraham et al. (2004). N2\_68 was observed in long-slit mode with
the ISIS spectrograph at the William Herschel Telescope in a similar
manner to the observations described in Willott et al. (2003). An
optical spectrum of N2\_64 was also obtained with ISIS after the
near-infrared spectrum failed to reveal a redshift.

\subsection{Redshifts}
\label{redshifts}

The five OHS near-IR spectra are shown in Fig.\,\ref{fig:spec}.
Emission lines are detected in four of the five X-ray sources. Two
sources (N2\_15 and N2\_60) have \halpha\,$\lambda6563$ and
\oiii\,$\lambda5007$ emission lines detected. The velocity offset of
+700 kms$^{-1}$ from \halpha\ to \oiii\ in N2\_15 is quite common in
high-redshift quasars (McIntosh et al. 1999). In N2\_47 there is one
strong emission line which we identify as \halpha\ and a probable
detection of \sii\,$\lambda6716/6731$ at the same redshift. N2\_61 has
just one secure emission line. We believe this is \halpha\ due to its
high equivalent width and velocity width of $5000$\,km\,s$^{-1}$
identifying it as a permitted line. There is a spike in the spectrum
at the location of \oiii\,$\lambda5007$ for this redshift, but it is
not definitely a real feature. N2\_64 shows no emission lines in the OHS
spectrum.

The optical spectrum of N2\_68 showed a blue object with two strong,
broad ($2000-3000\, {\rm km\,s}^{-1}$) emission lines. The wavelengths
of these lines indicate they are \lya\ and \civ\ at a redshift of
$z=2.218$. A one-dimensional spectrum could only be extracted from the
blue-arm data; there was insufficient signal in the red-arm to locate
the quasar. The spectrum is shown and used along with photometry in
Sec.\,\ref{specfit} to constrain the amount of reddening of this
quasar.

The optical spectrum of N2\_35 has a blue continuum and several
emission lines. The strongest emission line at 5790\,\AA\ has a deep
absorption trough close to its center and what appears to be an
absorption trough a few thousand km/s blueward of the peak. Other
weak, marginally significant emission lines are visible at 6139\,\AA\
and 7142\,\AA (these look more significant in the 2D spectrum than in
the 1D spectrum). The locations of these emission lines gives us a high
degree of certainty that the redshift is $z=2.741$ and that the
strongest line is \civ\ rather than \lya\ at $z=3.76$. The tentative
absorption trough blueward of \civ\ could indicate this to be a broad
absorption line quasar. The spectrum is shown and discussed in
Sec.\,\ref{specfit}.

Only N2\_64 does not show any reliable emission lines in either its
optical or near-IR spectra. The optical spectrum does show weak
continuum extending down to 3800\,\AA\ without a break, suggesting a
redshift of $z<2.1$. We note that for redshifts in the ranges
$1.0<z<1.3$ or $1.7<z<2.0$ the two strongest lines in the spectra of
N2\_15, N2\_47 and N2\_60 would not be observed in the wavelength
range covered by the OHS spectra, so these are likely redshift ranges
for N2\_61. The lack of a \lya\ emission line in the optical spectrum
supports the $1.0<z<1.3$ range, but reddened quasars can have very
weak \lya\ lines due to resonant dust scattering (Charlot \& Fall
1991) so the higher range is also possible. We have considered the
possibility that N2\_64 is a BL Lac object with optical and X-ray
fluxes dominated by beamed synchrotron. However, this object is not
detected in a very deep 1.4\,GHz radio observation (rms $9\mu$Jy;
Ivison et al. 2003) which leads us to conclude that it is not a BL
Lac.

\section{Spectral fitting}
\label{specfit}
Single power-law continuum fits to the OHS spectra were attempted for
all five quasars with near-IR spectra. Wavelength regions containing
quasar emission lines were masked out and an iterative chi-squared fit
was performed. The best-fit power-law slopes, $\alpha_\nu$, and the
associated reduced chi-squared, $\chi^2_{\nu}$, are labelled on
Fig.\,\ref{fig:spec}. The mean optical spectral index of radio-quiet
quasars is $\alpha_\nu=0.4$ (Brotherton et al. 2001). Two of the
quasars, N2\_15 and N2\_61 have very steep spectra with $\alpha_\nu
\geq 2$. The other three quasars have $0.5< \alpha_\nu < 0.8$, which
is slightly steeper than the mean spectral index but within the red
tail of the distribution observed in optically-selected samples such
as the SDSS (Richards et al. 2003). For N2\_15, N2\_60 and N2\_64 a
power-law provides a good description of the continuum shape
($\chi^2_{\nu} \approx 1$). N2\_47 has $\chi^2_{\nu} \approx 2$ and
shows deviations from a power-law, particularly a dip at a rest-frame
wavelength of $\approx 0.8 \umu$m which is also apparent in the SDSS
composite quasar spectrum (Vanden Berk et al. 2001). Although for
N2\_61 the best fit has $\chi^2_{\nu} \approx 1.5$, the fitted slope
of $ \alpha_\nu=2.01$ describes the overall continuum slope well.

Table 1 lists the optical $V$, $R$ and $I$ and near-infrared $H$
magnitudes of these quasars. At the redshifts of these quasars, the
observed optical emission is probing the rest-frame ultraviolet. The
colors of a typical unreddened quasar would be $V-I \approx 0.5$ and
$V-H \approx 2$. It is clear from this table that the rest-frame UV
and UV--optical colors of the quasars we have observed are
considerably redder than those of a typical quasar. 


\begin{figure}
\vspace{0.1cm}
\hspace{0.05cm}
\resizebox{0.48\textwidth}{!}{\includegraphics{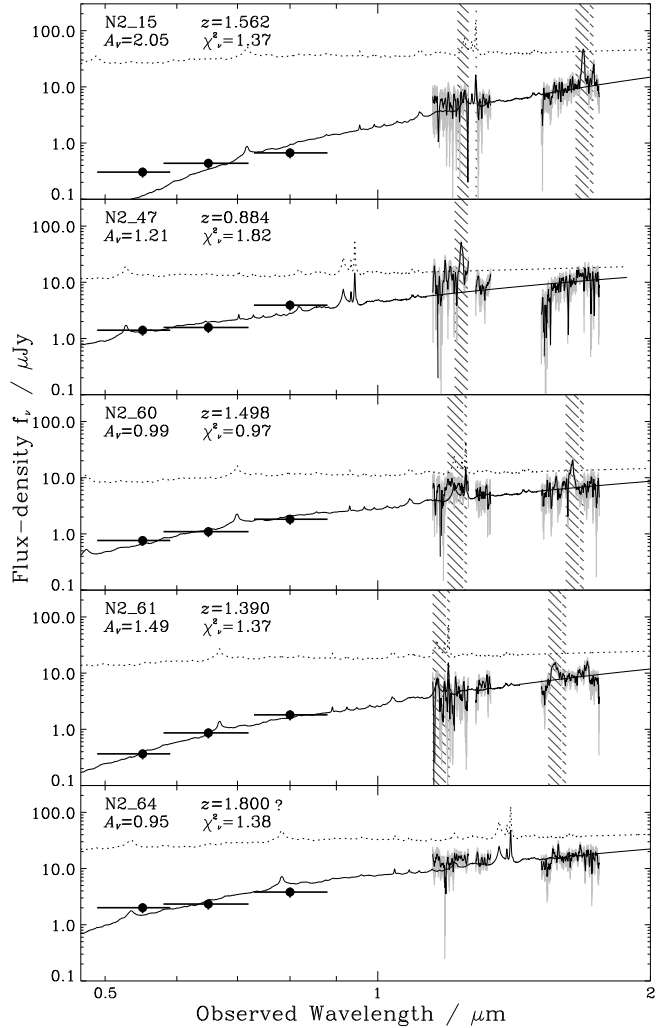}}
\vspace{0.3cm}
\caption{Observed-frame optical photometry and near-infrared spectra of the five
reddened quasars observed with OHS.  Near-IR spectroscopy is shown as
the solid line surrounded by grey shading to indicate the $\pm 1
\sigma$ noise. Optical photometry in the $V$, $R$ and $I$ bands are
shown with filled circles at the effective wavelengths of the
passbands with horizontal lines showing the extent of the passbands
and a vertical line the photometric errors. The best-fit reddened
composite quasar spectra are shown (assuming a SMC extinction law
only). The values of the extinction $A_V$ and the goodness-of-fit
$\chi^2_{\nu}$ refer to this fit. The dotted quasar spectrum is the
intrinsic, unreddened spectrum. Regions of the near-IR spectra
excluded from the fitting due to the presence of emission lines are
hatched.
\label{fig:photspec}}
\end{figure}

The red colors of these quasars can be explained by an intrinsic shape
similar to ordinary quasars which is reddened by a foreground screen
of dust. There are other possible explanations for red quasars: (i) an
optical-synchrotron dominated spectrum (e.g. Francis et al. 2001);
(ii) a host galaxy dominated spectrum (e.g. Vanden Berk et
al. 2001). We discount the former possibility because optical
synchrotron only dominates in powerful radio-loud quasars with flat
radio spectra and the latter because our near-infrared observations
show that these sources are unresolved in 0.5 arcsec seeing.

To constrain the amount of reddening required to produce these red
quasars we have performed a fit to the optical photometry and
near-infrared spectroscopy. The fact that our objects are spatially
unresolved in both the optical and near-IR means we can use a single
component quasar fit. For a typical quasar template we use the
composite radio-quiet quasar spectrum from the FIRST Bright Quasar
Survey (FBQS; Brotherton et al. 2001). We use this rather than the
SDSS composite of Vanden Berk et al. (2001) because the SDSS composite
shows a break at $>4000$\,\AA\ where contamination by host galaxy
light becomes important. In contrast, the FBQS radio-quiet quasar
composite spectrum shows a constant power-law from the UV right out to
6000\,\AA. We extend the composite redward of 6000\,\AA\ by continuing
the power-law slope of $\alpha_\nu=0.43$. We do not add in emission
lines such as \halpha, since we will only be using the line-free
regions of the quasar spectrum for continuum fitting. The composite
quasar spectrum is then subjected to a variable amount of dust
reddening and an iterative $\chi^2$ routine determines the best-fit
amount of reddening. The routine allows for reddening with three
different extinction laws: that of our galaxy, the LMC and the SMC,
adopting the parametric extinction laws of Pei et al. (1992).

The optical photometry and near-infrared spectroscopy for these five
quasars are shown in Fig.\,\ref{fig:photspec}. Also plotted are the
results of the reddened quasar fitting process - both the reddened
spectrum and the original unreddened spectrum. This illustrates that
none of the five quasars have optical and near-infrared spectra
similar to the unreddened quasar template. It also shows that the
observed flux in the optical has been heavily affected by absorption
(by factors of $10-100$). Although the fitting was performed for three
different extinction laws, only the SMC extinction law is shown in
this plot. The $\chi^2_{\nu}$ values for the different extinction laws
varied by $<10$\% for most objects (the main exception being N2\_15
where a Galactic extinction law fits the optical photometry much
better and has a $\chi^{2}_{\nu}=1.1$). The best-fit $A_V$ values are
systematically higher for LMC (120\% of SMC) and Galactic (130\% of
SMC) type dust because of their lower ratios of UV to optical
absorption. With the limited data available we are not able to draw
any firm conclusions on which dust extinction law provides the best fit to
these reddened quasars.

\begin{figure}
\vspace{0.1cm}
\vspace{-0.2cm}
\resizebox{0.48\textwidth}{!}{\includegraphics{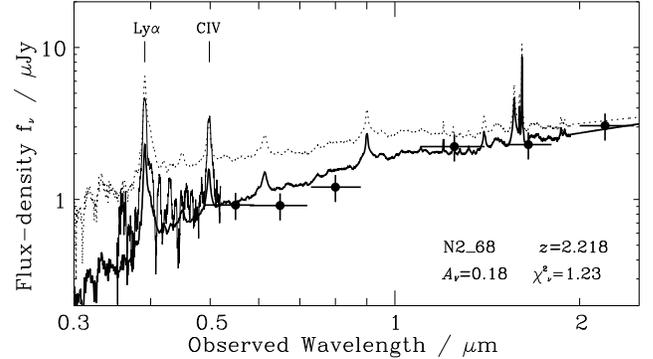}}
\vspace{-0.4cm}
\caption{ISIS blue-arm optical spectrum ($0.35 - 0.52\umu$m) and optical
($V,R,I$) and near-IR ($J,H,K$) photometry points for N2\_68.  The
best-fit reddened composite quasar spectrum is shown with a thick line
(assuming a SMC extinction law only). The dotted quasar spectrum is
the intrinsic, unreddened spectrum. The best-fit quasar has a small
amount of reddening ($A_V\approx 0.2$), but it is possible that the
near-IR photometry is dominated by host galaxy light, in which case
the optical data can be well fit by a quasar with no reddening.
Regions of the optical spectrum containing the \lya\ and \civ\
emission lines and data shortward of \lya\ are excluded from the
fitting.
\label{fig:n268}}
\end{figure}

\begin{figure}
\vspace{-0.2cm}
\resizebox{0.48\textwidth}{!}{\includegraphics{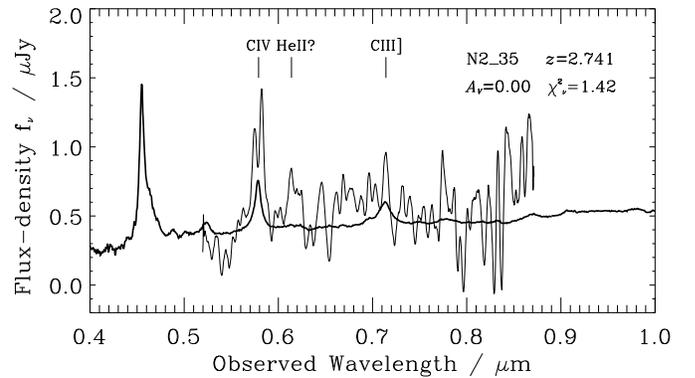}}
\vspace{-0.4cm}
\caption{Gemini-N GMOS optical spectrum (thin line) of N2\_35.  The
best-fit composite quasar spectrum is shown with a thick line. The
best-fit quasar has no reddening. Regions of the optical spectrum
containing the labelled emission lines are excluded from the fitting.
\label{fig:n235}}
\end{figure}


For N2\_68, we do not have near-IR spectroscopy but we do have an
optical spectrum from the blue-arm of ISIS at the WHT and optical
($V,R,I$) and near-IR ($J,H,K$) photometry. The snr in the optical
photometry is sufficient to tell that the source is spatially
unresolved. However, the source is faint in the near-IR ($J=22.14
\pm0.11$, $H=21.65\pm0.16$, $K=20.83\pm0.18$) and it is not possible
to tell if the near-IR emission is resolved or not. A passively evolving
$L_{\star}$ galaxy formed in an instantaneous starburst at $z=5$ has
$K=20.2$ at the redshift of N2\_68, so it is certainly possible that
the near-IR fluxes are dominated by extended, host galaxy emission. In
Fig.\,\ref{fig:n268} we show the optical spectrum and photometry data
for N2\_68. We have performed reddened quasar fitting to these data in
a manner similar to described previously for the other EDXS
quasars. We find there is relatively little reddening for N2\_68
($A_V\approx 0.2$ for an SMC extinction law and $A_V\approx 0.4$ for a
Milky Way law). If the near-IR fluxes are dominated by the host
galaxy, then the data are consistent with an unreddened quasar.

N2\_35 does not have near-IR spectroscopy but does have an optical
spectrum described in Sec.\,\ref{redshifts} and shown in
Fig.\,\ref{fig:n235}. This spectrum is used to fit the same range of
reddened quasar models as detailed above. N2\_35 is very faint in the
near-IR and is undetected at the $J$ and $H$ bands. It is detected at
$K$-band with $K=21.04\pm0.16$ and is clearly resolved in that image
(0.6 arcsec seeing). As mentioned in the preceding paragraph, this
magnitude is consistent with that of a quasar host galaxy at
this redshift.  Therefore the $K$-band photometry was not included in
the quasar fitting algorithm. The optical spectrum appears to show no
reddening. Note that the {\em quasar} $K$-band magnitude predicted by
this fit is $K=22.4$ which explains why the host galaxy outshines the
quasar at $K$.

\section{Comparison of X-ray and optical absorption}
\label{xoptabs}


\begin{figure*}
\hspace{0.6cm}
\resizebox{0.96\textwidth}{!}{\includegraphics{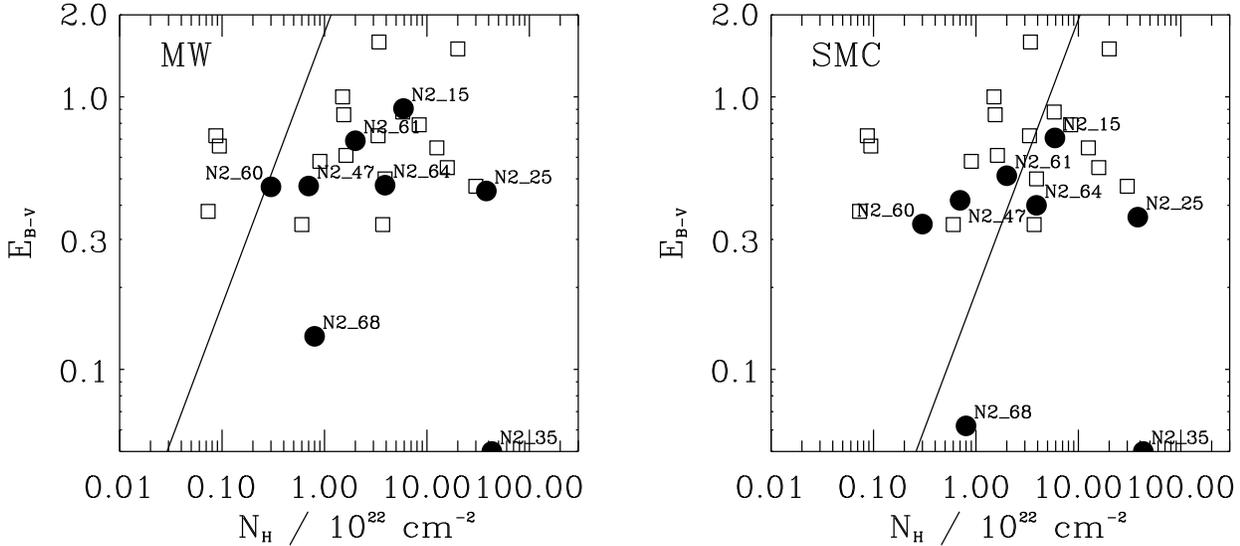}}
\caption{A comparison of the optical reddening due to dust $E(B-V)$
and gas absorption column density $N_{\rm H}$ for the optically-faint
EDXS-N2 quasars (circles). {\it Left:} The dust extinction values
plotted on this panel are for the fits with a Milky-Way extinction
law. The curve shows the standard Galactic ratio. Also plotted are the
Seyfert galaxies studied by Maiolino et al. (2001a; open
squares). {\it Right:} The dust extinction values plotted here are for
the fits with a SMC extinction law and the curve shows the gas-to-dust
ratio for the SMC. The Seyfert data are plotted in this panel with the
same values of $E(B-V)$ as for the Milky Way, since the reddening was
derived from observations at wavelengths greater than 4000 \AA\, where
the two extinction curves are the same. N2\_35 shows no measurable
optical extinction but is plotted at the minimum $E(B-V)$ on these
plots.
\label{fig:avnh}}
\end{figure*}


For eight of the nine quasars in our optical magnitude limited EDXS-N2
sample, we have constrained the amount of reddening of the optical
quasar emission. The optical dust reddening $E(B-V)$ is calculated
from the fitted $A_V$ values assuming the standard values of $R_V=3.1$
(Galactic) or $R_V=2.9$ (SMC). We can determine the gaseous X-ray
absorbing column by using the observed \chandra hardness ratio and
assuming an intrinsic power-law with $\Gamma=2$ as found by Mainieri
et al. (2002). X-ray absorption in the hard, soft and full bands is
determined following the prescription of Morrison and McCammon (1983).

Fig.\,\ref{fig:avnh} plots the optical reddening due to dust against
the absorbing column due to gas for these quasars. The left panel
shows the results of fits with Galactic dust and the right panel is
for SMC dust. Also shown are curves for a Galactic gas-to-dust ratio
($N_H=5.8 \times 10^{21} E(B-V)$; Bohlin et al. 1978) and a SMC
gas-to-dust ratio ($N_H=5.2 \times 10^{22} E(B-V)$; Bouchet et
al. 1985). The factor of 10 difference between these is attributed to
the lower metallicity in the SMC meaning less metals available for
dust formation.

The EDXS quasars tend to have higher gas-to-dust absorption ratios
than Galactic, although some of them are quite close to the curve and
do not show a statistically significant deviation from the Galactic
value. In contrast, the quasars are evenly distributed around the SMC
gas-to-dust ratio. Note that sources with $E(B-V)>1$ are not likely to
be in our sample because their rest-frame UV emission may be too
highly obscured for them to pass the $R<25$ magnitude limit and/or the
host galaxy may dominate giving an extended appearance. Similarly,
there are many quasars with $E(B-V)<0.3$ which have $R<23$ and
were therefore not followed up in this paper (P\'erez-Fournon et
al. 2004). So this diagram should be interpreted as the gas column
densities of a nearly complete sample of lightly reddened quasars. 

The discovery of N2\_25 with a gas-to-dust absorption ratio 100 times
Galactic (Willott et al. 2003) along with other reports of high
gas-to-dust ratio in the literature were the motivation for this
study. The only other quasar with such an extreme ratio is
N2\_35. This quasar appears to be completely unobscured in the
optical, but has a very hard X-ray spectrum. The optical spectrum
shows hints that this is a broad absorption line quasar - these
quasars are known to have hard X-ray spectra due to a large column
density (Gallagher et al. 2002), but comparatively little optical
reddening (Reichard et al. 2003). Apart from these two quasars with
extreme gas-to-dust absorption ratios, a ratio in the range 1-10 times
Galactic is typical of the rest of the sample.

Maiolino et al. (2001a) compared the gas and dust absorption
properties of a sample of nearby Seyfert galaxies and found that they
typically have higher ratios than Galactic. The data from this study
are also plotted on Fig.\,\ref{fig:avnh}. The Seyferts and EDXS
quasars tend to occupy similar locations on this plot, suggesting that
higher gas-to-dust absorption ratios in AGN are not a strong function
of luminosity and/or redshift. One should note that the particular
selection criteria used by Maiolino et al. (2001a) for their Seyfert
sample almost inevitably led to the sample being dominated by objects
with high gas-to-dust absorption ratios and the effect of any such
biases should become clearer with a volume-limited Seyfert sample
being studied with \xmm (Cappi et al. 2003).

The similarity of the gas-to-dust ratios in AGN and the SMC is not
easily explained. In the case of the SMC, the gas-to-dust ratio is
about 10 times higher than that of the galaxy because its metallicity
is about 10 times lower (Bouchet et al. 1985). In the case of AGN,
super-solar metallicities are common in the circumnuclear region
(Hamann \& Ferland 1999). In any case, metallicity effects are not so
important when the gas absorbing column density comes from an X-ray
measurement (Maiolino et al. 2001b). This is because the measured
X-ray absorption is due to photoelectric absorption by metals, so the
derived gas density is proportional to the metallicity. Given the
quite different physical conditions in AGN environments and in the
SMC, the similarity of these gas-to-dust ratios is likely just a
coincidence.

There are three basic ways to interpret the higher gas-to-dust
absorption ratios in the AGN environment as compared to the Milky Way:
(i) the dust grain composition is different so dust is less effective
at absorbing UV radiation; (ii) the majority of the X-ray absorption
occurs close to the ionizing radiation source, within the dust
sublimation radius; (iii) the ratio of gas to dust is higher than in
our galaxy. These possibilities are discussed in more detail in
Maiolino et al. (2001a; 2001b) and work still needs to be done to
determine which effect is dominant.

\begin{figure}
\resizebox{0.48\textwidth}{!}{\includegraphics{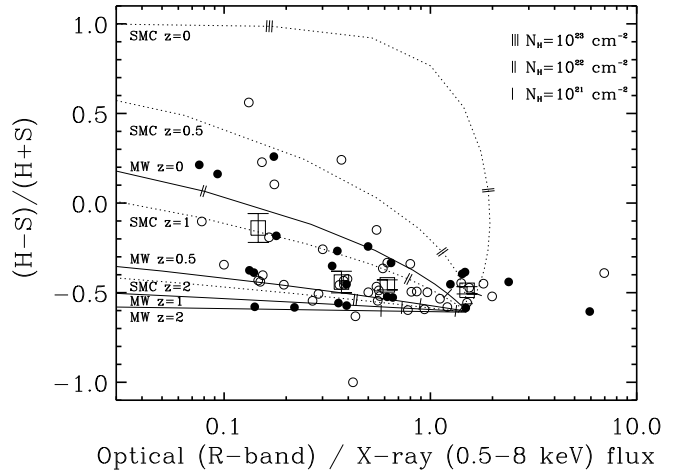}}
\vspace{0.05cm}
\caption{Same data as plotted in lower panel of Fig.\,\ref{fig:hard}.
The curves show how the observed hardness ratio of a quasar with
intrinsic optical to X-ray flux ratio of 1.5 and X-ray power-law of
slope $\Gamma=2.0$ would depend upon the observed optical to X-ray
flux ratio for a range of obscuring gas and dust properties at various
redshifts. At each of the four plotted redshifts a curve is shown for
Milky Way (solid) and SMC (dotted) dust (and the associated
gas-to-dust ratio) as described in Sec.\,\ref{xoptabs}. Each curve
corresponds to a range of absorbing columns which are indicated by
tickmarks along these curves.
\label{fig:hardmod}}
\end{figure}

We now return to the issue of the effect that obscuration has on the
observed optical to X-ray flux ratio. The lower panel of Fig.
\ref{fig:hard} and the statistics presented in Sec.\,\ref{hardratio}
showed that as the X-ray emission from quasars gets more heavily
obscured, the ratio of optical to X-ray flux decreases. Can this
obscuration be simply explained by an absorber composed of dust and
gas? To test this we have simulated the effects of gas and dust on the
optical and X-ray flux of quasars at various redshifts. The intrinsic
X-ray slope of the quasars are modeled with a $\Gamma=2$ power-law.
The same procedure as detailed above was used to relate $N_H$, $HR$
and the amount of full-band X-ray absorption.  The optical absorption
in the observed frame $R$-band was calculated by assuming the relevant
extinction curve of Pei et al. (1992). Again, we consider both the
Milky Way and SMC extinction laws and their gas-to-dust ratios. We now
replot the data from Fig.\,\ref{fig:hard} on Fig.\,\ref{fig:hardmod}
adding in curves which show how the optical to X-ray flux ratio and
hardness change as one increases the column of absorbing material. We
plot Galactic and SMC curves for four different redshifts.  Since the
optical flux is much more readily depressed than the full band X-ray
flux for typical gas-to-dust ratios, the optical to X-ray flux ratio
decreases as the absorbing column is increased (except for small
columns at $z=0$ for the SMC). The strong redshift dependence of these
curves is due to the fact that the k-corrections for obscured AGN in
the optical/UV and X-rays have opposite trends.

Fig.\,\ref{fig:hardmod} shows that if the quasars were to lie mostly in
the redshift range $0<z<0.5$ then their range of hardness ratio and
optical to X-ray ratio could be explained by a small intrinsic scatter
coupled with absorption by gas and dust with Galactic
properties. However, very few of these quasars lie at such low
redshifts and spectroscopy (Sec.\,\ref{obs}; Barger et al. 2002) shows
that most of them lie within the redshift range $1<z<2.5$. At these
redshifts, simple Galactic gas and dust obscuration cannot
simultaneously explain the hardness ratios and optical to X-ray
ratios. In comparison, the curves for the SMC provide a much better
match to the data. This is mostly due to the 10 times higher
gas-to-dust ratio and not differences in the dust extinction laws.

In Willott et al. (2003) we speculated that if there are a large
fraction of AGN which are Type 1 in the optical (i.e. broad permitted
lines) and Type 2 in the X-ray (hard X-ray spectrum due to absorption)
then they could be a major contributor to the hard X-ray
background. This would lessen the need for optically-obscured sources
to dominate the hard X-ray background. We can obtain an estimate of
the contribution of such sources to the X-ray background from the
number of sources with unresolved optical counterparts and high
hardness ratios. From consideration of the hardness ratios and
Fig.\,\ref{fig:avnh} we specify sources with HR~$>-0.2$ as hard enough
to be Type 2 in the X-ray. This corresponds to an absorption column
density of $3\times 10^{22}~{\rm cm}^{-2}$ at $z=2$ assuming an
intrinsic $\Gamma=2$ power-law. In the EDXS-N2 region there are four
sources (N2\_15, N2\_25, N2\_35 and N2\_37) with HR~$>-0.2$ and
optically-unresolved. This is out of a parent sample of 61 sources
above the flux-limit of $>2 \times 10^{-15}$\,erg\,cm$^{-2}$\,s$^{-1}$
and with snr~$>5\sigma$, giving 6\% of the sample. A similar analysis
in the CDF-N also finds that 6\% of the sample have HR~$>-0.2$ and are
optically-unresolved. Since the sources in these samples are in the
flux range which dominates the hard X-ray background (50-64\% of the
2-8 keV background is resolved by the EDXS; Manners et al. 2003) we
conclude that only about 6\% of the sources responsible for the 2-8
keV X-ray background appear as Type 1 in the optical and Type 2 in the
X-ray.

\section{Effect of obscuration on observed optical to X-ray ratio}


\begin{figure}
\resizebox{0.48\textwidth}{!}{\includegraphics{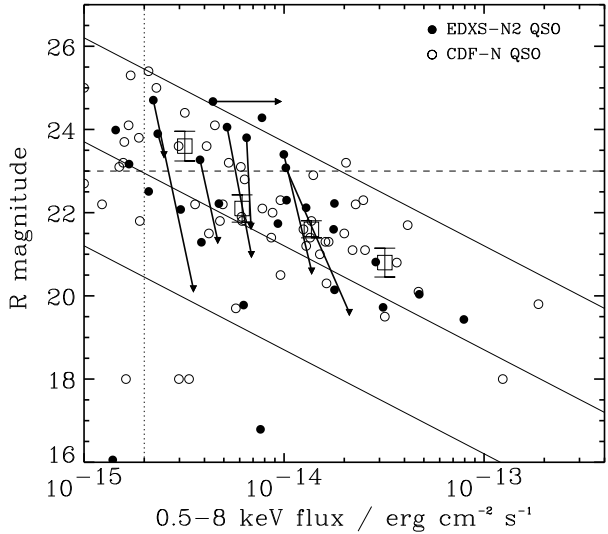}}
\caption{As in Fig.\,\ref{fig:optxflux} the $R$ magnitudes of EDXS-N2
(filled circles) and CDF-N (open circles) quasars are plotted against
full band X-ray flux. The squares with error bars are the median $R$
magnitude against median X-ray flux as in Fig.\,\ref{fig:optxflux}.
The arrows show the de-absorbed $R$ magnitudes and X-ray fluxes of the
EDXS-N2 quasars with $R>23$ for which the optical reddening has been
determined. The de-reddened $R$ magnitudes of these optically-faint
quasars brings them back into the typical range of optical to X-ray
flux ratio.
\label{fig:optxfluxdered}}
\end{figure}


Having shown that obscuration plays an important role in the observed
properties of Type 1 AGN, we now revisit the effect this may have on
the observed optical to X-ray ratio. Given the fact that at the
redshifts where most quasars are observed ($z>1$) the optical to X-ray
ratio is so strongly dependent upon the obscuration, this may skew the
observed distribution of optical to X-ray ratios away from their
intrinsic values.


Fig.\,\ref{fig:optxfluxdered} plots optical magnitude against
full-band X-ray flux for quasars from the EDXS-N2 and CDF-N
surveys. For eight of the nine EDXS-N2 quasars with observed
magnitudes $R>23$ we have determined the optical reddening and X-ray
absorption. We use here the reddening values determined with the SMC
extinction law, but note that the values are very similar with the
Milky Way law. Plotted on Fig.\,\ref{fig:optxfluxdered} are arrows
showing the locations of these quasars once they have been deabsorbed
in the optical and X-ray. The effect of absorption is quite dramatic,
particularly at optical wavelengths. This illustrates that these
quasars which are observed to be optically-faint are mostly not
intrinsically optically-faint. The main exception is N2\_35 which does
not show any optical absorption, but is heavily obscured in the X-rays
and hence deabsorbing it actually decreases the optical to X-ray flux
ratio

\begin{figure}
\vspace{0.9cm}
\resizebox{0.48\textwidth}{!}{\includegraphics{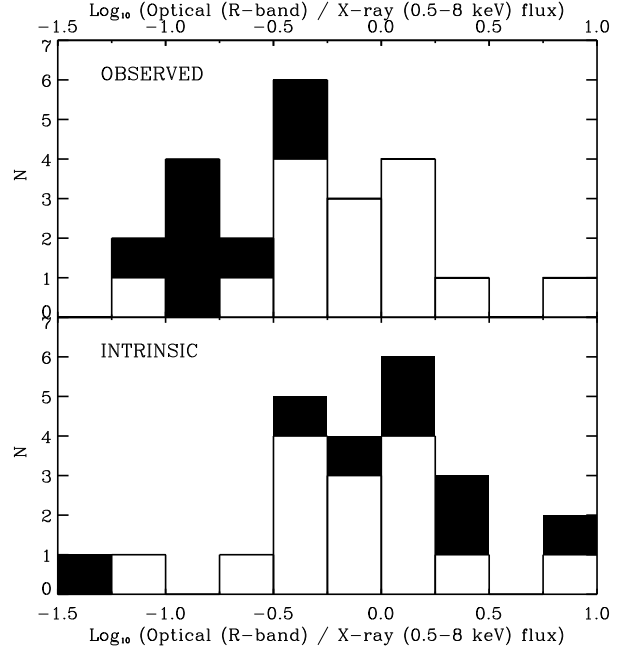}}
\vspace{0.4cm}
\caption{Histogram of optical to X-ray flux ratio for EDXS-N2
quasars. The upper panel shows the histogram of ratios as
observed. The black filled histogram shows the eight sources for which
the optical reddening has been determined. In the lower panel the
absorption-corrected optical to X-ray flux ratios for these eight
sources are shown in black. The effect of these corrections is to
shift the histogram to higher optical to X-ray flux ratios.
\label{fig:histdered}}
\end{figure}


To determine how big an effect this has on the whole population of
quasars, in Fig.\,\ref{fig:histdered} we show a histogram of the
optical to X-ray flux ratio for all the EDXS-N2 quasars with 0.5-8 keV
fluxes $>2 \times 10^{-15}$\,erg\,cm$^{-2}$\,s$^{-1}$. The upper panel
shows the observed optical to X-ray flux ratios. The lower panel shows
the intrinsic ratios for the sources shown with deabsorbing arrows in
Fig.\,\ref{fig:optxfluxdered}. Note that the absorption-corrected
histogram could be incomplete for two reasons. Firstly, there may be
quasars with $R>25$ and low values of optical-to-X-ray flux ratio that
are not identified as such due to the limited depth and resolution of
the optical imaging. Secondly, some of the quasars with $R<23$ will be
subject to obscuration and correcting for this would likely shift
their optical-to-X-ray flux ratios to higher values. Despite these
incompletenesses, this plot clearly shows that the typical intrinsic
value of the optical to X-ray flux ratio is greater in all but one
case (N2\_35 -- the possible broad absorption line quasar) than one
would think from looking at the raw data.

The ratio of optical and X-ray flux is often expressed in terms of
the optical to X-ray spectral index, $\alpha_{\rm ox}$, where 
\begin{equation}
\alpha_{\rm ox}=\frac{\log(f_{\rm 2~keV}/f_{2500~\mbox{\rm \scriptsize\AA}})}{\log(\nu_{\rm 2~keV}/\nu_{2500~\mbox{\rm \scriptsize\AA}})}.
\end{equation}
The $\alpha_{\rm ox}$ distribution for bright, low-redshift
optically-selected quasars comprises a roughly Gaussian distribution
with a peak at $\alpha_{\rm ox}=-1.5$ and a tail of about 10\% of the
quasars at $\alpha_{\rm ox}<-2$ (Brandt, Laor \& Wills 2000; Vignali
et al. 2003). The existence of broad or associated absorption lines in
most of these $\alpha_{\rm ox}<-2$ quasars suggests that soft X-ray
absorption is the reason for the higher optical to X-ray flux ratios
in these quasars. Anderson et al. (2003) have recently confirmed that
there is a weak anti-correlation between optical luminosity and
$\alpha_{\rm ox}$ and showed that for moderate luminosity X-ray
selected quasars a typical value of $\alpha_{\rm ox}$ is
$-1.4$. 

Two of the quasars in the EDXS-N2 sample (N2\_25 and N2\_35) are
absorbed by columns greater than $10^{23}~{\rm cm}^{-2}$. For these
quasars, the rest-frame 2 keV is below the photo-electric cut-off
leading to almost total absorption of the 2 keV flux. The calculated
observed $\alpha_{\rm ox}$ for these quasars assuming an X-ray
spectrum with absorption columns implied by the hardness ratios are
$\alpha_{\rm ox}=-3.3 ~{\rm and} -4.2$, respectively. Applying
corrections for the obscuration in the X-rays and optical (for N2\_25
only, since N2\_35 is optically-unobscured) leads to intrinsic values
of $\alpha_{\rm ox}=-1.0 ~{\rm and} -1.2$, respectively. The quasar
N2\_68 which shows little or no reddening or X-ray absorption has
$\alpha_{\rm ox} =-1.6$.

For the other five EDXS-N2 quasars discussed in this paper the observed
optical to X-ray spectral indices lie in the small range
$-1.3<\alpha_{\rm ox}<-1.0$. After corrections for optical and X-ray
absorption these spectral indices all get steeper and lie in the range
$-1.3<\alpha_{\rm ox}<-1.6$ which is typical of lower redshift quasars
with similar optical luminosities (Anderson et al. 2003). This is
interesting because the behaviour for high-redshift, X-ray-selected
quasars contrasts with that for low-redshift, optically-selected
quasars. In the optically-selected sample the obscuration is strongest
at X-ray energies, since strong reddening of the optical continuum
would push the optical magnitude above the magnitude limit. In the
X-ray-selected sample the obscuration is strongest in the
(observed-frame) optical. Note that the faint optical and X-ray flux
limits in our survey are such that a much more representative range of
absorbed quasars are present than in samples with brighter flux
limits.

Risaliti et al. (2001) showed that optical emission line selected
quasars tend to have steeper $\alpha_{\rm ox}$ than optical color
selected quasars. They interpret this as due to a larger amount of
X-ray absorption in the slightly reddened quasars and suggest this
shows that a large population of quasars exist which are Type 1 in the
optical and Type 2 in the X-rays. This behaviour is seen for N2\_25
and N2\_35 but not for the rest of our sample. As discussed in
Sec.\,\ref{xoptabs} we find that such quasars can only contribute a
small fraction of the X-ray background. The differences between our
findings and those of Risaliti et al. are likely due to two issues:
(i) their definition of the X-ray flux was based on soft X-rays (ROSAT
PSPC counts) which are more readily absorbed than the \chandra
full-band flux; (ii) their emission line selected quasars have a very
bright optical magnitude limit unlike ours which are X-ray selected
with faint optical magnitudes.

Under the assumption that most of the optically-bright ($R<23$)
EDXS-N2 quasars have little or no reddening, then their intrinsic
$\alpha_{\rm ox}$ values are similar to the absorption-corrected
values we have found. Thus the intrinsic spread in $\alpha_{\rm ox}$
is much narrower than the observed range. This is very important for
constraining the optical and X-ray radiation production
mechanisms. Since the EDXS-N2 sample does not yet have complete
redshift and reddening information, an examination of the intrinsic
spread in $\alpha_{\rm ox}$ is beyond the scope of this current paper,
but is clearly an important goal for the future.

\section{Conclusions}

Deep X-ray surveys are capable of uncovering the AGN responsible for
most of the supermassive black hole accretion history of the universe.
Combined X-ray and optical/near-infrared observations of lightly
reddened quasars from such X-ray surveys are a powerful probe of the
physical conditions of the obscuring material. We have used optical
and near-infrared photometry and spectroscopy to constrain the amount
of reddening in the light from optically-faint quasars. Our main
conclusions are:

\begin{itemize}

\item{Optically-faint ($R>23$) quasars at $f_{0.5-8}>2 \times
10^{-15}$\,erg\,cm$^{-2}$\,s$^{-1}$ are mostly faint due to
obscuration. De-reddening the observed $R$-band fluxes of these
quasars gives them optical magnitudes similar to other quasars with
these X-ray fluxes ($20<R<22$).}

\item{Assuming that most of the spread in X-ray hardness ratios is due
to absorption by gas, we find that gas-to-dust ratios in reddened
X-ray selected quasars are typically a few times greater than the
gas-to-dust ratio in the Milky Way and similar to those in Seyfert
galaxies.}

\item{The fraction of quasars responsible for the X-ray background
which are Type 1 in the optical (broad permitted lines) and Type 2 in
the X-ray (hard X-ray spectrum due to absorption) is only about 6\%.}

\item{The observed distribution of optical-to-X-ray flux ratios of
quasars at $z>1$ is skewed to low values compared to the intrinsic
distribution due to the fact that the observed-frame $R$-band light is
emitted in the UV and is more easily obscured than the hard X-rays
sampled by {\it Chandra}.}

\end{itemize}


\acknowledgments

We thank the anonymous referee for comments that improved this paper.
Based on data collected at Subaru Telescope, which is operated by the
National Astronomical Observatory of Japan. Also based on observations
obtained at the Gemini Observatory, which is operated by the
Association of Universities for Research in Astronomy, Inc., under a
cooperative agreement with the NSF on behalf of the Gemini
partnership: the National Science Foundation (United States), the
Particle Physics and Astronomy Research Council (United Kingdom), the
National Research Council (Canada), CONICYT (Chile), the Australian
Research Council (Australia), CNPq (Brazil) and CONICET (Argentina).
The William Herschel Telescope is operated on the island of La Palma
by the Isaac Newton Group in the Spanish Observatorio del Roque de los
Muchachos of the Instituto de Astrofisica de Canarias. Thanks to Steve
Rawlings for assistance with some of the optical spectroscopy. CJW
thanks the NRC for support. CS acknowledges the award of a PPARC
Advanced Fellowship.

\end{document}